\documentclass[12pt]{article}
\usepackage{amsmath,amssymb,amsthm,cite,enumerate}

\newtheorem{theorem}{Theorem}
\newtheorem{corollary}{Corollary}
{\theoremstyle{definition}
\newtheorem{remark}{Remark}
}

\begin{document}
\begin{center}
\par\noindent {\LARGE\bf
Group analysis of variable coefficient\\ generalized fifth-order KdV equations
\par}

{\vspace{5mm}\par\noindent\large Oksana Kuriksha$^{\dag 1}$, Severin Po{\v s}ta$^{\ddag 2}$ and Olena~Vaneeva$^{\S 3}$
\par\vspace{2mm}\par}
\end{center}

{\par\noindent\it\small
${}^\dag$\ Petro Mohyla Black Sea State University, 10, 68 Desantnykiv Street, 54003 Mykolaiv, Ukraine\\[1ex]
${}^\ddag$\ Department of Mathematics, Faculty of Nuclear Sciences and Physical Engineering,\\[1ex]
 $\phantom{{}^\ddag}$\ Czech Technical University in Prague, 13 Trojanova Str., 120 00 Prague, Czech Republic \\[1ex]
${}^\S$\ Institute of Mathematics of  NAS of Ukraine, 3 Tereshchenkivska Str., 01601 Kyiv-4, Ukraine
}

{\vspace{3mm}\par\noindent
$\phantom{{}^\dag{}\;}\ $E-mails: \it $^1$oksana.kuriksha@gmail.com, $^2$severin.posta@fjfi.cvut.cz,
$^3$vaneeva@imath.kiev.ua
\par}

{\vspace{5mm}\par\noindent\hspace*{5mm}\parbox{150mm}{\small

We carry out group analysis of a class of  generalized fifth-order Korteweg-de Vries equations with time dependent coefficients.
Admissible transformations, Lie symmetries and similarity reductions of equations from the class  are classified exhaustively. A criterion of reducibility  of variable coefficient fifth-order KdV equations to their constant coefficient
counterparts is derived. Some exact solutions are presented.

}\par\vspace{4mm}}

\section{Introduction}
The classical Korteweg-de Vries  (KdV) equation and its generalizations model various  physi\-cal systems, including
gravity waves, plasma waves and waves in lattices~\cite{jeffrey}. In particular, the KdV equation arises
in the modeling of one-dimensional plane waves in cold quasi-neutral collision-free plasma
propagating along the $x$-direction
under the presence of a uniform magnetic field~\cite{Kakutani1968}. It appeared that,
when the propagation angle of the wave relative to the external magnetic
field becomes critical, the third-order (dispersion) term in the model equation should be replaced by the fifth-order one~\cite{Kakutani&Ono1969}. Namely, magneto-acoustic waves propagating along this critical direction are modeled by
the simplest fifth-order KdV (fKdV) equation (called also quintic KdV equation),
\begin{gather}\label{eq_fKdV}
u_t+uu_x+\mu u_{xxxxx}=0, \quad \mu=\mathrm{const}.
 \end{gather}
In \cite{Nagashima1979} the equation~\eqref{eq_fKdV} with $\mu=-1$ was shown to describe solitary waves in the nonlinear transmission line of a LC
ladder type.

Later equation \eqref{eq_fKdV} and its generalizations were  studied in a number of papers. Thus, an exact solitary wave solution of equation~\eqref{eq_fKdV} in terms of Jacobi elliptic function $\mathrm{cn}$ was found in \cite{Kano&Nakayama1981,yamamoto}.
Pulsating multiplet solutions of this equation
were examined in \cite{hyman}. Local conservation laws with the densities $u$, $u^2$ and $u^3+\frac32(u_{xx})^2$ were indicated therein.
 Note that the fKdV equation is not integrable by the inverse scattering
transform method in contrast to the classical KdV equation~\cite{MikhailovShabatSokolov1991}. Lie symmetries and the corresponding reductions of  \eqref{eq_fKdV} to ordinary differential equations (ODEs) were found in~\cite{Liu2010f}.

In the last decades there is a great interest to variable coefficient models that in many cases describe the real world phenomena with more accuracy. Classifications of Lie symmetries are usual tasks in studies of such models. This is due to the fact
that Lie symmetries allows one not only to reduce a model PDE to a PDE with fewer number of independent variables or to an ODE but
also  to derive cases that are  potentially more interesting for applications~\cite{FN}.

 An attempt of Lie symmetry classification of the generalized fKdV equations with time dependent coefficients, $u_t+u^nu_x+\alpha(t)u+\beta(t)u_{xxxxx}=0$, was made in~\cite{Wang&Liu&Zhang2013}. However, the results presented therein are incorrect in general.
In the present paper we perform the correct and complete group classification of the class
\begin{gather}\label{eq_ggfKdV}
u_t+uu_x+\alpha(t)u+\beta(t)u_{xxxxx}=0,\quad \beta\neq0,
\end{gather}
where $\alpha$ and $\beta$ are smooth  functions
of the variable $t$.
To be able to reduce the number of variable coefficients and to proceed with Lie symmetry analysis in an optimal way, we at first find the admissible transformations~\cite{popo2010a} (called also allowed~\cite{Winternitz92} or form-preserving~\cite{Kingston&Sophocleous1998} ones)
in class~\eqref{eq_ggfKdV}. Classifications of Lie symmetries and similarity reductions are presented in Sections~3 and 4, respectively.

\section{Admissible transformations}Roughly speaking an admissible transformation is a triple consisting  of two fixed equations from a class
and a~point transformation linking these equations. The set of admissible  transformations of a class of DEs possesses the groupoid structure with respect to the standard composition of transformations~\cite{Popovych&Bihlo2012}. More details and examples on finding and utilizing  admissible transformations for fKdV-like equations as well as definitions of different kinds of equivalence groups are given in~\cite{Kuriksha&Posta&Vaneeva2014JPA,VPS2013}.

To classify
admissible transformations in class~\eqref{eq_ggfKdV} we
suppose that an equation from~\eqref{eq_ggfKdV} is connected with an equation from the same class,
\begin{equation}\label{eq_ggKawahara_tilda}
\tilde u_{\tilde t}+\tilde u\tilde u_{\tilde x}+\tilde \alpha(\tilde t) \tilde u+\tilde \beta(\tilde t) \tilde u_{\tilde x\tilde x\tilde x\tilde x\tilde x}=0,
\end{equation}
via a nondegenerate point~transformation in the space of variables $(t,x,u)$. Without loss of generality the consideration can be restricted
to point transformations
of the special form
\begin{gather}\label{EqEquivtransOfvcKdVlikeSuperclass}
\tilde t=T(t),\quad \tilde x=X^1(t)x+X^0(t),\quad \tilde u=U^1(t,x)u+U^0(t,x),
\end{gather}
where $T$, $X^i$,  and $U^i$, $i=0,1$, are arbitrary smooth functions of their variables with $T_tX^1U^1\neq0$.
This restriction is true for any subclass of the class of evolution equations of the form
$
u_t=F(t)u_n+G(t,x,u,u_1,\dots,u_{n-1}),$ with $F\ne0,$ and $G_{u_iu_{n-1}}=0,$ $ i=1,\dots,n-1.
$
Here $n\geqslant 2$,
$
u_{n}=\frac{\partial^n u}{\partial x^n},
$
$F$ and $G$ are arbitrary smooth functions of their variables~\cite{VPS2013}.
We make the change of variables~\eqref{EqEquivtransOfvcKdVlikeSuperclass} in~\eqref{eq_ggKawahara_tilda}
and further substitute $u_t=-uu_x-\alpha(t)u-\beta(t)u_{xxxxx}$ to the obtained equation
in order to confine it to the manifold defined by~\eqref{eq_ggfKdV} in the fifth-order jet space
with the independent variables $(t,x)$ and the dependent variable~$u$.
Splitting the obtained identity with respect to the derivatives of $u$ leads to the determining
equations for the functions~$T$, $X^1$, $X^0$, $U^1$ and $U^0$. After certain simplifications the equations become
\begin{gather*}
\tilde \beta T_t=\beta(X^1)^5,\quad U^1_x=0,\quad U^1T_t=X^1,\quad U^0T_t=X^1_tx+X^0_t,\\
U^0_t=-\tilde\alpha T_tU^0,\quad U^1_tX^1-\alpha U^1 X^1+U^1U^0_x T_t + \tilde\alpha U^1 T_t X^1=0.
\end{gather*}
We solve these equations and get the following assertion.

\begin{theorem} The generalized extended equivalence group~$\hat G^{\sim}$ of  class~\eqref{eq_ggfKdV}
 is formed by the transformations
\begin{gather*}
\tilde t=T(t),\quad \tilde x=(x+\delta_1)X^1+\delta_2,\quad
\tilde u=\frac{1}{T_t}\left( X^1 u +X^1_t(x+\delta_1)\right), \\[1ex]
\tilde \alpha(\tilde t)=\frac{1}{T_t}\left(\alpha(t)-2\frac{X^1_t}{X^1}+\frac{T_{tt}}{T_t}\right),\quad\tilde\beta(\tilde t)=\dfrac{(X^1)^5}{T_t}\beta(t),
\end{gather*}
where $X^1=(\delta_3\int\! e^{-\int \alpha(t) {\rm d}t}{\rm d}t+\delta_4)^{-1},$ $\delta_j,$ $j=1,\dots,4,$ are arbitrary constants with
$(\delta_3,\delta_4)\not=(0,0)$ and $T=T(t)$ is a smooth function with $T_t\neq0$.

The entire set of admissible transformations of class~\eqref{eq_ggfKdV} is
generated by the transformations from the group~$\hat G^{\sim}$.
\end{theorem}
Using this theorem we can formulate a criterion of reducibility of variable coefficient fKdV equations to constant coefficient ones.
\begin{theorem}
A variable coefficient equation from class~\eqref{eq_ggfKdV} is reducible to the constant coefficient fKdV equation~\eqref{eq_fKdV} if and only if its coefficients $\alpha$ and $\beta$ are related by the formula
\begin{gather}\label{criterion2}
\beta=e^{-\int \alpha(t) {\rm d}t}\left(c_1\int\! e^{-\int \alpha(t) {\rm d}t}{\rm d}t+c_2\right)^3,
\end{gather}
where $c_1$ and $c_2$ are arbitrary constants with $(c_1,c_2)\not=(0,0).$
\end{theorem}

Using the equivalence
transformation
\begin{equation}\label{eq_gauge}
\hat t=\int e^{-\int \alpha(t)\, {\rm d}t}{\rm d}t,\quad \hat x=x, \quad
\hat u=e^{\int \alpha(t)\, {\rm d}t}u
\end{equation}
from the group~$\hat G^{\sim}$ we can set
the arbitrary element~$\alpha$ to the zero value.
Indeed, this transformation
maps
class~\eqref{eq_ggfKdV}  to its subclass  with $\hat
\alpha=0$. The arbitrary element $\hat\beta$ of a mapped equation is expressed in terms of $\alpha$ and $\beta$ as $\hat\beta=e^{\int \alpha(t)\, dt}\beta$.
Without loss of generality we can restrict ourselves to the investigation of the class
\begin{gather}\label{eq_gfKdV}
u_t+uu_x+\beta(t)u_{xxxxx}=0
\end{gather}
since all results on symmetries, classical solutions, conservation laws  and other related objects for equations from class~\eqref{eq_ggfKdV} can be found
using the similar results obtained for equations from class~\eqref{eq_gfKdV}.

We derive equivalence transformations in class~\eqref{eq_gfKdV}   setting $\tilde\alpha=\alpha=0$ in transformations presented in Theorem~1.
\begin{corollary}
The usual equivalence group~$G^\sim_{\alpha=0}$ of class~\eqref{eq_gfKdV}
consists of the transformations
\begin{gather*}
\tilde t=\dfrac{at+b}{ct+d},\quad
\tilde x=\dfrac{e_2x+e_1t+e_0}{ct+d},\quad
\tilde u=\dfrac{e_2(ct+d)u-e_2cx-e_0c+e_1d}\Delta,\quad
\tilde \beta=\dfrac{e_2{}^5}{(ct+d)^3}\dfrac {\beta}\Delta,
\end{gather*}
where $a$, $b$, $c$, $d$, $e_0$, $e_1$ and $e_2$ are arbitrary constants with $\Delta=ad-bc\ne0$ and $e_2\ne0$,
the tuple  $(a,b,c,d,e_0,e_1,e_2)$ is defined up to a nonzero multiplier
and hence without loss of generality we can assume that $\Delta=\pm1$.

The entire set of admissible transformations of class~\eqref{eq_gfKdV} is
generated by the transformations from the group~$G^\sim_{\alpha=0}$.
\end{corollary}
The transformation components for $t,$ $x$ and $u$ coincide with those obtained for the class of Burgers equations $u_t+uu_x+\beta(t)u_{xx}=0$~\cite{PP2012}
and the class of KdV equations $u_t+uu_x+\beta(t)u_{xxx}=0$~\cite{Popovych&Vaneeva2010}.
\begin{corollary}
A variable coefficient equation from class~\eqref{eq_gfKdV}  is reducible to the constant coefficient fKdV equation~\eqref{eq_fKdV} if and only if
$
\beta=(c_1t+c_2)^3,
$
where $c_1$ and $c_2$ are arbitrary constants with $(c_1,c_2)\not=(0,0).$

\end{corollary}

\section{Lie symmetries}
 To perform the group classification of class~\eqref{eq_gfKdV}
we use the  classical technique~\cite{Ovsiannikov1982}.
Namely, we look for
Lie symmetry operators of the form $Q=\tau(t,x,u)\partial_t+\xi(t,x,u)\partial_x+\eta(t,x,u)\partial_u$ that generate one-parametric Lie groups
of transformations leaving equations from class~\eqref{eq_gfKdV} invariant. The
Lie invariance criterion is written as
\begin{equation}\label{c2}
Q^{(5)}(u_t+uu_x+\beta(t)u_{xxxxx})\big|_{u_t=-uu_x-\beta(t)u_{xxxxx}}=0,
\end{equation}
where  $Q^{(5)}$ is the fifth prolongation of the operator~$Q$~\cite{Olver1986,Ovsiannikov1982}.
Equation~\eqref{c2} leads to the determining equations for the coefficients $\tau$, $\xi$ and $\eta$, the simplest of which result in
\[
\tau=\tau(t),\quad
\xi=\xi(t,x), \quad
\eta=\eta^1(t,x)u+\eta^0(t,x),
\]
where $\tau$, $\xi$, $\eta^1$ and $\eta^0$ are arbitrary smooth functions of their variables.
The rest of the determining equations have the form
\begin{gather*}
\tau \beta_t =(5\xi_x-\tau_t)\beta,\quad \eta^1_x=2\xi_{xx},\quad \eta^1_{xx}=\xi_{xxx},\quad 2\eta^1_{xxx}=\xi_{xxxx},\\
\eta^1_xu^2+(\eta^0_x +\eta^1_t+\eta^1_{xxxxx}\beta)u+\eta^0_t+\eta^0_{xxxxx}\beta=0,\\
(\tau_t-\xi_x+n\eta^1)u+(5\eta^1_{xxxx}-\xi_{xxxxx})\beta-\xi_t+\eta^0=0.
\end{gather*}
The latter two equations can be split with respect to the variable $u$.  After splitting we solve those equations that do not involve arbitrary element $\beta$ and get the general  form of the infinitesimal generator,
\[
Q=(c_2t^2+c_1t+c_0)\partial_t\!+\!((c_2t+c_3)x+c_4t+c_5)\partial_x\!+\!((c_3-c_1 -c_2t)u+c_2x+c_4)\partial_u,
\]
where $c_i$, $i=0,\dots5$, are arbitrary constants.
The single classifying equation is
\begin{equation*}
(c_2t^2+c_1t+c_0)\beta_t=(3c_2t-c_1+5c_3)\beta.
\end{equation*}
If the arbitrary element $\beta$ varies, then we can split the latter equation with respect to $\beta$ and its derivative $\beta_t$. As a result, we obtain that
$c_0=c_1=c_2=c_3=0$ and
the kernel $A^{\rm ker}$ of the maximal Lie invariance algebras of equations from class~\eqref{eq_gfKdV}
coincides with the two-dimensional algebra $\langle\partial_x,\,t\partial_x+\partial_u\rangle$.
To exhaustively describe cases of Lie symmetry extension we should integrate the classifying equation with respect to $\beta$ up to $G^\sim_{\alpha=0}$-equiva\-lence.
Since the procedure is quite similar to that of the Lie symmetry classification for KdV equations, $u_t+uu_x+\beta(t)u_{xxx}=0$, we omit  details of calculations and refer the interested reader to~\cite{Popovych&Vaneeva2010}. The following assertion  is true.
\begin{table}[t!]\footnotesize \renewcommand{\arraystretch}{1.8}
\begin{center}\label{TableLieSym}
\textbf{Table~1.}
The group classification of the class~$u_t+uu_x+\beta(t) u_{xxxxx}=0$.
\\[2ex]
\begin{tabular}{|c|c|l|}
\hline
&$\beta(t)$&\hfil Basis of $A^{\max}$ \\
\hline
0&$\forall $&
$\partial_x,\,t\partial_x+\partial_u$\\
\hline
1&
$t^\rho$&$\partial_x,\,t\partial_x+\partial_u,\,5t\partial_t+(\rho+1) x\partial_x+(\rho-4) u\partial_u$\\
\hline
2&$e^{t}$&
$\partial_x,\,t\partial_x+\partial_u,\,5\partial_t+x\partial_x+u\partial_u$\\
\hline
3&$(t^2+1)^\frac32 e^{5\nu\arctan t}$&
$\partial_x,\,t\partial_x+\partial_u,\,(t^2+1)\partial_t+(t+\nu)x\partial_x+((\nu-t)u+x)\partial_u$\\
\hline
4&$1$&
$\partial_x,\,t\partial_x+\partial_u,\,\partial_t,\,5t\partial_t+x\partial_x-4u\partial_u$\\
\hline
\end{tabular}
\\[2ex]
\parbox{155mm}{Here   $\rho$ and $\nu$ are real constants, $\rho\neq0$. Up to $G^\sim_{\alpha=0}$-equivalence we can assume that $\rho\leqslant3/2$, $\nu\geqslant0$.}\\[3ex]
\textbf{Table~2.}
The group classification of class~$u_t+uu_x+\alpha(t)u+\beta(t)u_{xxxxx}=0$ using no equivalence.
\\[2ex]
\begin{tabular}{|c|c|l|}
\hline
no.&$\beta(t)$&\hfil Basis of $A^{\max}$ \\\hline
0&$\forall$&$\partial_x,\ T\partial_x+T_t\partial_u$\\ \hline
1&$\lambda T_t(aT+b)^\rho(cT+d)^{3-\rho}$
&$ \partial_x,\ T\partial_x+T_t\partial_u,\  5T_t^{-1}(aT+b)(cT+d)\partial_t+\bigl[5acT+$\\
&&$ad(\rho+1)+bc(4-\rho)\bigr]x\partial_x+\Bigl(5acxT_t-\bigl[5acT+$\\
&&$5\alpha T_t^{-1}(aT+b)(cT+d)+bc(\rho+1)+ad(4-\rho)\bigr]u\Bigr)\partial_u$\\ \hline
2&$\lambda T_t(cT+d)^3\exp\left(\frac{aT+b}{cT+d}\right)$&
$ \partial_x,\ T\partial_x+T_t\partial_u,\  5T_t^{-1}(cT+d)^2\partial_t+(5c(cT+d)+\Delta)x\partial_x+$\\
&&$\left[5c^2xT_t+\left(\Delta-5(cT+d)(c+\alpha(cT+d) T_t^{-1})\right)u\right]\partial_u$\\ \hline
3&$\lambda T_te^{5\nu\arctan\left(\frac{aT+b}{cT+d}\right)}\mathcal{H}^3$&
$ \partial_x,\ T\partial_x+T_t\partial_u,\ T_t^{-1}\mathcal{H}^2\partial_t+\bigl[a(aT+b)+c(cT+d)+\nu\Delta\bigr] x\partial_{ x}$\\
&&$+\bigl[(a^2+c^2)xT_t-\bigl(a(aT+b)+c(cT+d)-\nu\Delta+\alpha T_t^{-1}\mathcal{H}^2\bigr)u\bigr]\partial_u$\\
 \hline
4a&$\lambda T_t$&
$ \partial_x,\ T\partial_x+T_t\partial_u,\   T_t^{-1}(\partial_t-\alpha u\partial_u),\, 5TT_t^{-1}\partial_t+x\partial_x-(4+5TT_t^{-1}\alpha)u\partial_u$\\ \hline
4b&$T_t(cT+d)^3$&
$ \partial_x,\ T\partial_x+T_t\partial_u,\,5T_t^{-1}(cT+d)\partial_t+4cx\partial_x-(c+5T_t^{-1}(cT+d)\alpha)u\partial_u,$\\
&&$T_t^{-1}(cT+d)^2\partial_t+c(cT+d)x\partial_x+$\\
&&$[c^2xT_t-(cT+d)(c+T_t^{-1}(cT+d)\alpha)u]\partial_u$\\\hline
\end{tabular}
\\[2ex]
\parbox{155mm}{Here $a$, $b$, $c$, $d$, $\lambda$, $\nu$, and $\rho$ are arbitrary constants with $\lambda\neq0$ and $\rho\neq0,3$, $\Delta=ad-bc\ne0$; $\mathcal{H}=\sqrt{(aT+b)^2+(cT+d)^2}.$ The function $\alpha(t)$ is arbitrary in all cases, $T=\int e^{-\int \alpha(t) {\rm d}t}{\rm d}t$. }
\end{center}
\end{table}
\begin{theorem}
The kernel of the maximal Lie invariance algebras of equations from class~\eqref{eq_gfKdV}
is the two-dimensional Abelian algebra $A^{\rm ker}=\langle\partial_x,\,t\partial_x+\partial_u\rangle$.
All possible $G^\sim_{\alpha=0}$-inequiva\-lent  cases of extension of the maximal Lie invariance algebras are exhausted
by Cases $1$--$\,4$ of Table~1.
\end{theorem}
\begin{remark}
A group classification list for class~\eqref{eq_ggfKdV} up to $\hat G^\sim$-equivalence coincides with the list presented in Table~1.
\end{remark}
\begin{remark}
An equation of the form~\eqref{eq_ggfKdV} admits a four-dimensional Lie symmetry algebra if and only if it is point-equivalent to the constant coefficient fKdV equation~\eqref{eq_fKdV}.
\end{remark}

In Table 2 we present also the complete list of Lie symmetry extensions for  class~\eqref{eq_ggfKdV},
where arbitrary elements are not simplified by point transformations. This is achieved using the equivalence-based approach~\cite{Vaneeva2012}.

Cases presented in Table~2  give all equations~\eqref{eq_ggfKdV} for which
the classical method of Lie reduction can be effectively used.

\section{Lie symmetry reductions}
Lie symmetries provide one with
the powerful tool for finding solutions of nonlinear PDEs reducing them to PDEs with fewer number of independent variables or even to ODEs.
If a (1+1)-dimensional PDE admits a Lie symmetry operator,
 $Q=\tau\partial_t+\xi\partial_x+\eta\partial_u$,
then the ansatz reducing this PDE to an ODE is found as a solution of the invariant surface condition $Q[u]:=\tau u_t+\xi u_x-\eta=0$~\cite{Olver1986,Ovsiannikov1982}. In practice,  one has to solve the corresponding characteristic system $\frac{{\rm d}t}{\tau}=\frac{{\rm d}x}{\xi}=\frac{{\rm d}u}{\eta}$. To get inequivalent
reductions one should use subalgebras from an optimal system (see Section~3.3 in~\cite{Olver1986}).

We have constructed optimal systems of one-dimensional subalgebras for all the maximal Lie invariance algebras presented in Table~1.
The results are summarized in Table~3.
\begin{table}[h!]\footnotesize\renewcommand{\arraystretch}{1.8}
\begin{center}
\textbf{Table 3.} Optimal systems of one-dimensional subalgebras of $A^{\rm max}$  presented in Table 1.
\\[2ex]
\begin{tabular}
{|c|l|}
\hline
\hfil Case
&
\hfil Optimal system
\\
\hline
0
&
${\mathfrak g}=\langle\partial_x\rangle,
\quad
{\mathfrak g}^a=\langle (t+a)\partial_x+\partial_u\rangle$
\\
\hline
$1_{\rho\neq-1,4}$
&
${\mathfrak g}=\langle\partial_x\rangle,
\quad
{\mathfrak g}^{\sigma}=\langle (t+\sigma)\partial_x+\partial_u\rangle,\quad {\mathfrak g}_{1.1}=\langle 5t\partial_t+(\rho+1) x\partial_x+(\rho-4) u\partial_u\rangle$
\\
\hline
$1_{\rho=-1}$
&
$\mathfrak g=\langle\partial_x\rangle,
\quad
\mathfrak g^{\sigma}=\langle(t+\sigma)\partial_x+\partial_u\rangle,
\quad
\mathfrak g^a_{1.2}=\langle t\partial_t+a\partial_x-u\partial_u\rangle$ \\
\hline
2
&
$\mathfrak g=\langle\partial_x\rangle,
\quad
\mathfrak g^0=\langle t\partial_x+\partial_u\rangle,
\quad
\mathfrak g^{\,}_{2}=\langle5\partial_t+x\partial_x+u\partial_u\rangle$ \\
\hline
3
&
$\mathfrak g=\langle\partial_x\rangle,
\quad
\mathfrak g^{\,}_{3}=\langle(t^2+1)\partial_t+(t+\nu)x\partial_x+(x+(\nu-t)u)\partial_u\rangle$
\\
\hline
4
&
$\mathfrak g=\langle\partial_x\rangle,
\quad
\mathfrak g^{\,}_{4.1}=\langle\partial_t\rangle,\quad\mathfrak g^\sigma_{4.2}=\langle \sigma\partial_t+t\partial_x+\partial_u\rangle$,\quad $\mathfrak g_{4.3}=\langle 5t\partial_t+x\partial_x-4u\partial_u\rangle$
\\
\hline
\end{tabular}
\end{center}
\parbox{155mm}{Here $a$ is a real constant,   $\sigma\in\{-1,0,1\}$. Up to $G^\sim_{\alpha=0}$-equivalence we can assume that $\rho\leqslant3/2$, $\nu\geqslant0$.}
\end{table}

The reductions with respect to the subalgebra $\mathfrak g$ lead to constant solutions only.
The reduction with respect to the subalgebra $\mathfrak g_{4.3}$ is not presented since it coincides with that performed using $\mathfrak g_{1.1}$ for $\rho=0$.
Other reductions are listed in Table~4.

\begin{table}[h!]\footnotesize \renewcommand{\arraystretch}{1.8}
\begin{center}
\textbf{Table 4.} Similarity reductions of the equations~$u_t+uu_x+\beta(t)u_{xxxxx}=0$.
\\[2ex]
\begin{tabular}
{|c|c|c|c|l|}
\hline
Case&$\mathfrak g$& $\omega$ &\hfil Ansatz, $u=$ &\hfil Reduced ODE
\\
\hline
0 &${\mathfrak g}^a$& $t$ & $\varphi(\omega)+\dfrac{x}{t+a}$ & $(\omega+a)\varphi'+\varphi=0$
\\
\hline
$1_{\rho\ne-1,4}$&${\mathfrak g}^{\,}_{1.1}$&$xt^{-\frac{\rho+1}{5}}$ &
$t^{\frac{\rho-4}{5}}\varphi(\omega)$ & $ \varphi'''''+ \left(\varphi-\frac{\rho+1}{5}\omega\right)\varphi'+\frac{\rho-4}{5}\varphi=0$
\\
\hline
$1_{\rho=-1}$ & ${\mathfrak g}^a_{1.2}$&$x-a\ln t$ & $t^{-1}\varphi(\omega)$ & $ \varphi'''''+ \left(\varphi-a\right)\varphi'-\varphi=0$
\\
\hline
2 &${\mathfrak g}_2$& $xe^{-\frac15t}$ & $e^{\frac1{5}t}\varphi(\omega)$ & $
 \varphi'''''+\left(\varphi-\frac1{5}{\omega}\right)\varphi'+\frac1{5}\varphi=0$
\\
\hline
3 &${\mathfrak g}^{\,}_{3}$& $\dfrac{xe^{-\nu\arctan t}}{\sqrt{t^2+1}}$ &
$\dfrac{e^{\nu\arctan t}}{\sqrt{t^2+1}}\varphi(\omega)+\dfrac{xt}{t^2+1}$ & $ \varphi'''''+(\varphi-\nu\omega)\varphi'+\nu
\varphi+\omega=0$\\
\hline
4.1 &${\mathfrak g}_{4.1}$ &$x$ & $\varphi(\omega)$ & $
 \varphi'''''+\varphi\varphi'=0$
\\
\hline
4.2 &${\mathfrak g}^\sigma_{4.2}$& $x\pm\dfrac{t^2}2$ & $\varphi(\omega)\mp t$ & $ \varphi'''''+\varphi\varphi'\mp1=0$
\\[1.5mm]
\hline
\end{tabular}
\end{center}
\parbox{155mm}{Here $a$ is an arbitrary constant.}
\end{table}
{\samepage Solving the first-order reduced equation from Table~4 and subsequently applying to it transformation~\eqref{eq_gauge}
we get a ``degenerate'' solution of equation~\eqref{eq_ggfKdV}, \[u=\dfrac{x+b}{\int\! e^{-\int \alpha(t) {\rm d}t}{\rm d}t+a}\,e^{-\int \alpha(t){\rm d}t},\]
that is valid for any smooth function $\alpha$. Here $a$ and $b$ are arbitrary constants.}

Using equivalence transformations it is possible to construct an exact solution for the equations~\eqref{eq_ggfKdV} that are reducible to their constant
coefficient counterparts, i.e., whose coefficients are related by~\eqref{criterion2}. We take the known solution in terms of the Jacobi elliptic function~$\mathrm{cn}$ from~\cite{yamamoto} for equation~\eqref{eq_fKdV} and get the exact solution
\[u=\dfrac{\dfrac{105}{16}a\,\,\mathrm{cn}^4\!\left(\dfrac{\sqrt{2}}4a^\frac14\left(\dfrac{x+d}{Z}-\dfrac{21}8a\displaystyle\int \frac{e^{-\int \alpha(t) {\rm d}t}}{Z^2}\,{\rm d}t\right)+b;\dfrac{\sqrt{2}}2\right)+c_1(x+d)}{e^{\int \alpha(t) {\rm d}t}Z}\]
for the variable coefficient fKdV equation,
\[u_t+uu_x+\alpha(t)u-e^{-\int \alpha(t) {\rm d}t}Z^3u_{xxxxx}=0,\]
where $Z=c_1\int\! e^{-\int \alpha(t) {\rm d}t}{\rm d}t+c_2$, $a$ is a positive constant, $c_1$, $c_2$, $b$ and $d$ are arbitrary constants with $(c_1,c_2)\not=(0,0).$
\section{Conclusion}

In the present paper,  the group classification problem for
class~\eqref{eq_ggfKdV} of variable coefficient fKdV
equations, which appear in various gravity and plasma wave models, is completely solved.
The use of the generalized extended equivalence group~$\hat G^\sim$ has allowed us to present the
classification list in a rather simple  form (Table~1).
For  the sake of convenience in further applications, we also write down the classification list extended
by equivalence transformations (Table~2). The Lie symmetry algebra of an equation from class~\eqref{eq_ggfKdV} is of maximal dimension (which is equal to four) if this equation has  constant coefficients or is point-equivalent to one with constant coefficients.

\looseness=-1
One-dimensional subalgebras of the Lie symmetry algebras admitted by equations from class~\eqref{eq_ggfKdV} are classified in Table~3
and all inequivalent reductions with respect to such subalgebras are summarized in Table~4. Performed reductions can be used for the construction of exact and/or  numerical solutions. Examples of such constructions were given in~\cite{Kuriksha&Posta&Vaneeva2014JPA} for the generalized Kawahara equations. Two simple solutions are also constructed in the present paper.

\bigskip

\noindent
{\bf Acknowledgements.}
The authors would like to acknowledge partial support of their participation in SQS'13 kindly provided by the Organizers.
They are also grateful to Roman Popovych for useful comments.

\end{document}